# Pt/Ti/Al$_2$O$_3$/Al tunnel junctions showing electroforming-free bipolar resistive switching behavior

Doo Seok Jeong,[1,a)] Byung-ki Cheong[1], and Hermann Kohlstedt[2,b)]

[1]*Electronic Materials Center, Korea Institute of Science and Technology,*
*39-1 Hawolgok-dong, Seongbuk-ku, Seoul 136-791, Republic of Korea*
[2]*Nanoelektronik, Technische Fakultät, Christian-Albrechts-Universität zu Kiel, D-24143 Kiel, Germany*

We investigated electroforming-free bipolar resistive switching behavior in Pt/Ti/Al$_2$O$_3$/Al tunnel junctions where the Al$_2$O$_3$ tunnel barrier was naturally formed on Al in air. Various compliance current values for the junction's set switching successfully lead to various resistance values in its low resistance state, suggesting the possibility for multi-level-operation. A mechanism for the bipolar switching is qualitatively discussed in terms of the modulation of the tunnel barrier by the reactive Ti layer on top of the barrier.

a) dsjeong@kist.re.kr
b) hko@tf.uni-kiel.de

Recently, crossbar arrays are considered as an interesting approach for ultra-large-scaled Gbit resistive random access memories (RRAMs) circuits as well as neuromorphic logic circuits based on synapse-mimic analog switches comprising $TiO_2$ memristor.[1-3] Switching elements applicable to these devices need a pre-treatment process termed electroforming, which leads to the formation of nm-sized conduction filaments.[4] However, the formation and rupture of the filament is regarded to be rather stochastic in nature.[5] In subsequent resistance-programming cycles, large parameter spread between individual cells could be a critical obstacle for the application of resistive switching phenomena to electronic devices. Moreover, when electroforming is conducted on crossbar arrays, each cell of the arrays will be electroformed under different voltages due to the voltage drop along the word and bit lines of the arrays. This consequently results in a cell-to-cell variation of switching behavior.

The objective of the present investigation is to seek the answer to the question "How can we achieve an electroforming-free resistive switch on the basis of metal-insulator-metal (MIM) structure?". In more general terms, the idea has the following background: Even after a vast number of scientific publications, the nature of electroforming is not well understood. It is believed that thermal effects, hot-electron injection, charge trapping, ionic mass transport, and complicated inner electric field distribution are synchronized in electroforming. Probably due to the complicated nature of electroforming, it is hardly controllable by an external electric stimulus. A reasonable approach is therefore to avoid electroforming as an initial step for RRAM cells. The development of reproducible resistive switching devices based on non-filamentary behavior is therefore an interesting approach to overcome the current limitations of curiosity-driven filamentary switching devices.[6]

Tunnel junctions are versatile devices for diverse applications, for instance, tunnel diode, magnetic and ferroelectric tunnel junctions, and multi state memories.[7-12] Tunnel junctions with particularly engineered layer sequence may be the solution of the problems mentioned above. A key element in electron tunneling is the extreme sensitivity of tunneling current to the tunnel barrier properties (tunneling matrix element) as well as the electronic structure of the electrodes sandwiching the tunnel barrier (electron supply function). Interestingly, the first few unit cells of each electrode, which are in the vicinity of the metal/barrier interface, define the electron supply function. In case of an oxide tunnel barrier, by incorporating, for instance, an oxygen-reactive metal at one interface, an oxygen-ion-related electrochemical reaction can take place at the interface.[13] That is, the resistance of this tunnel junction is likely to be variable depending on the oxidation state of the contact metal. We term this tunnel junction a resistive tunnel junction (RTJ).

In this study, we chose native Al-oxide and Ti as a tunnel barrier and an oxygen-reactive contact metal in our RTJs, respectively. Al, Ti, and their oxides are widely used materials in the current semiconductor mass production. The layer sequence of the RTJs is illustrated in Fig. 1. By applying a sufficiently large external electric field, mobile oxygen ions might be attracted to or repelled from Al-oxide through the Al-oxide/Ti interface. By following the discussion above, this ion mass transport might lead to a variation of the tunnel junction current, accompanied with resistance switching in the current-voltage (*I-V*) characteristic. The native Al-oxide tunnel barrier most probably contains of a mixture of oxide ($Al_2O_3$) and hydroxide [$Al(OH)_3$] and its surface may be covered with the hydroxide since it was formed in air.[14,15] The coexistence of oxide and hydroxide was confirmed by an O1s x-ray photoelectron spectroscopy (XPS) spectrum of the tunnel barrier as shown in Fig. 1(c). The thickness of the tunnel barrier was obtained to be approximately 3 nm from an Al2p XPS spectrum shown in Fig. 1(b). Moreover, the Ti2p spectrum in Fig. 1(a) points out that the Ti layer was oxidized into various valence states, e.g., TiO, $Ti_2O_3$, and $TiO_2$. This oxidation may transform the underneath $Al(OH)_3$ into $Al_2O_3$ as does a Nb layer on $Al(OH)_3$.[14]

Pt/Ti/$Al_2O_3$/Al tunnel junctions were formed in crossbar structure. 100 nm thick Al bottom electrodes were patterned on an oxidized Si wafer using a lift-off process and deposited using electron-beam evaporation. By placing the wafer in air for approximately 2 hours a native $Al_2O_3$ tunnel barrier on the Al bottom electrode was formed. By using a subsequent lift-off process and electron-beam evaporation, a Pt/Ti bilayer top electrode was then deposited on the tunnel barrier to complete the junctions. The Pt/Ti thicknesses were 100 nm and 10 nm, respectively. The role of the upper Pt layer was to prevent the oxidation of the reactive Ti layer in air. Resistive switching measurement was performed by applying a staircase dc voltage sweep to the top electrode while the bottom electrode was grounded using a Keithely 236 Source Measure Unit. A measurement delay time of 1 msec. was set for the measurements.

Contrary to filament-driven switching, electroforming was not necessary to activate resistive switching in the fabricated RTJs. A typical *I-V* curve undergoing resistive switching is plotted in Fig. 2. In order to prevent dielectric breakdown of the tunnel barrier, a compliance current (cc) of 500 μA was set. In Fig. 2, it can be seen that set switching (high resistance state (HRS)→low resistance state (LRS)) and reset switching (LRS→HRS) occurs under negative and positive voltages, respectively. Since no electroforming is involved, the cell-to-cell uniformity in switching is excellent. Statistics of switching voltages obtained from 7 different junctions of each pad-size is plotted in Fig. 3(a), which shows almost negligible standard deviation. Furthermore, the HRS is found to be identical to the as-fabricated state as shown in Figs. 2 and 3(a). This results from the lack of electroforming. The set and the reset switching voltages shown in Fig. 3(a) were defined by the crossing point between the HRS and the LRS dI/dV-V branches on the negative and the positive voltage sides, respectively. A typical dI/dV vs.V curve is plotted in Fig. 3(b).

In Fig. 2, one can notice that both set and reset switching behaviors are gradual rather than abrupt as opposed to filament-driven switching. This aspect opens up the possibility of multi-level-operation as well as for analog-switching–operation of the RTJs under well-controlled operation conditions. We also performed switching measurement with different cc values. By setting cc, maximum current flow, as well as the applied voltage maximum could be controlled. Switching curves measured with six different cc values are plotted in Fig. 4. The applied voltage maximum of each cc case is depicted in the inset of Fig. 4. It was found that the LRS resistance gradually decreases with increasing cc (increasing maximum set voltage). The decrease in the LRS resistance by a factor of approximately 3 over the voltage range (1.2 – 1.7 V), as shown in the inset of Fig. 4, means that there is a sufficient voltage margin for multi-level-operation.

We have identified three possible mechanisms for the observed resistive switching, each related to a specific region of the junction: 1. The $Al_2O_3/Al$ interface, 2. The $Al_2O_2$ tunnel barrier interior, and 3. The $Ti/Al_2O_3$ interface. Starting with the $Al_2O_3/Al$ interface, this interface most probably moves towards the anode of the tunnel junction, implying that set and reset switching should take place under positive and negative voltage application to the top electrode. However this is not consistent with the observed switching polarity shown in Fig. 3. We can therefore rule out the $Al_2O_3/Al$ interface effects as a possible mechanism.

Secondly, self-diffusion of ions in the $Al_2O_3$ tunnel barrier under the influence of applied voltage is also possible. However, assuming that the migrated ions will undergo redistribution to their original state as soon as the applied voltage is removed is hard to accept as a mechanism for the observed non-volatile switching. Therefore, the $Ti/Al_2O_3$ interface is our only remaining candidate.

The conductivity of $TiO_x$ decreases as $x$ increases so that the electron supply function of $TiO_x$ varies depending on the oxidation state of Ti. When a negative voltage is applied to the Pt/Ti top electrode, the initially oxidized Ti in its as-fabricated state, as shown in Fig. 1(a), may be reduced by the applied over-potential. The reduction of the $TiO_x$ layer produces a higher electric field in the tunnel than the one before the reduction, which consequently leads to an increase in the conductivity of the RTJ (set switching). During the reset switching (positive voltage application to the Pt/Ti electrode), the reduced $TiO_x$ may be re-oxidized so that its conductivity decreases, which consequently leads to a decrease in an electric field in the tunnel barrier. Therefore, the conductivity of the RTJ may decrease.

Although we suggested a possible mechanism for the observed resistive switching, further elaboration of the mechanism, especially, the kinetics of the oxygen-ion-exchange electrochemical reaction is needed. Detailed roles of the electrochemical reaction in the change in the tunneling current also need to be investigated.

In summary, we have introduced $Pt/Ti/Al_2O_3/Al$ RTJs that show electroforming-free resistive switching behavior with excellent cell-to-cell uniformity. Furthermore, the non-abrupt set and reset switching behaviors of our devices have the advantage of multi-level-operation of RRAMs and even analog switches. We consider the $Ti/Al_2O_3$ interface to play a crucial role in the resistive switching, where redox reactions involving Ti are thought to take place, with the reactions influence the tunneling current through the tunnel barrier.

**Acknowledgements**

DSJ and BC would like to acknowledge a research grant by Korean Ministry of Education, Science, and Technology through an institutional research program of Korea Institute of Science and Technology and the Fundamental R&D Program for Core Technology of Materials funded by the Ministry of Knowledge Economy, Republic of Korea. We would like to thank Dr. Michael Hambe and Dr. Rohit Soni for carefully reading the manuscript. We also thank Prof. Cheol Seong Hwang for fruitful comments.

**Figure captions**

Figure 1. Stack sequence of the RTJ. XPS spectra of (a) Ti2p of the Ti layer, (b) Al2p and (c) O1s of the Al-oxide tunnel barrier. The spectra in (b) and (c) were taken before the deposition of the top electrode. The spectrum in (a) was taken after dry-etching the Pt layer of the fabricated stack.

Figure 2. Typical *I-V* switching curve of the Ti/$Al_2O_3$/Al RTJ and an *I-V* curve in an as-fabricated state. The inset is a schematic of the RTJ. The voltage was applied to the top electrode of the RTJ and the bottom electrode was grounded.

Figure 3. (a) Set and reset switching voltages with respect to various pad-sizes of the RTJ. The voltages were taken from switching curves of 7 different junctions of each pad-size in order to demonstrate cell-to-cell uniformity in switching behavior. The resistance ratio of the pristine state to the HRS is also plotted with respect to the pad-sizes. (b) *I-V* switching and dI/dV curves. The crossing point on each polarity side between the dI/dV-V branches belonging to the HRS and the LRS, respectively, is defined as a representative switching voltage.

Figure 4. *I-V* switching curves measured with various cc values. HRS and LRS resistances and maximum set voltage as a function of the cc values are plotted in the inset.

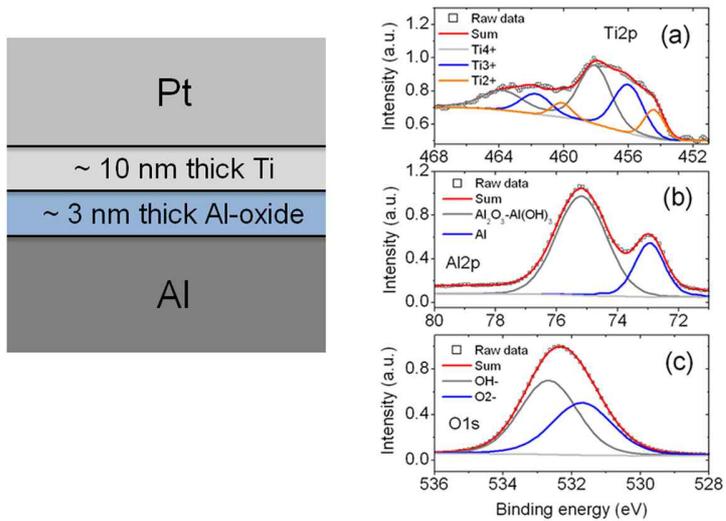

Figure 1

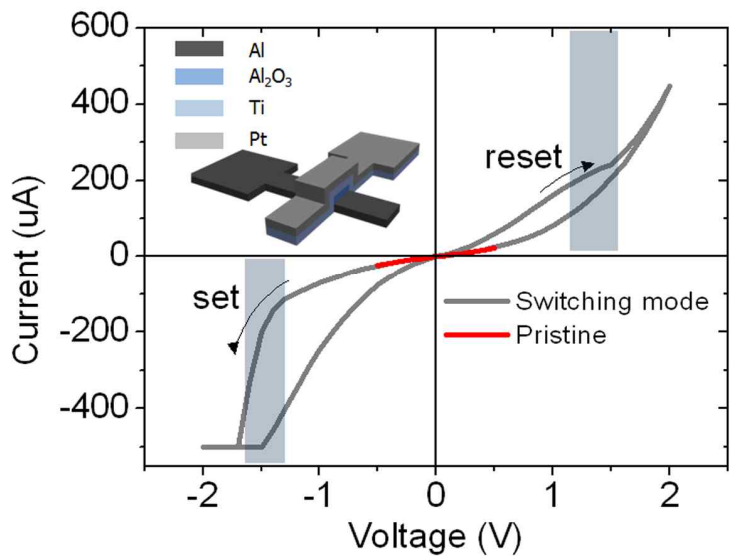

Figure 2

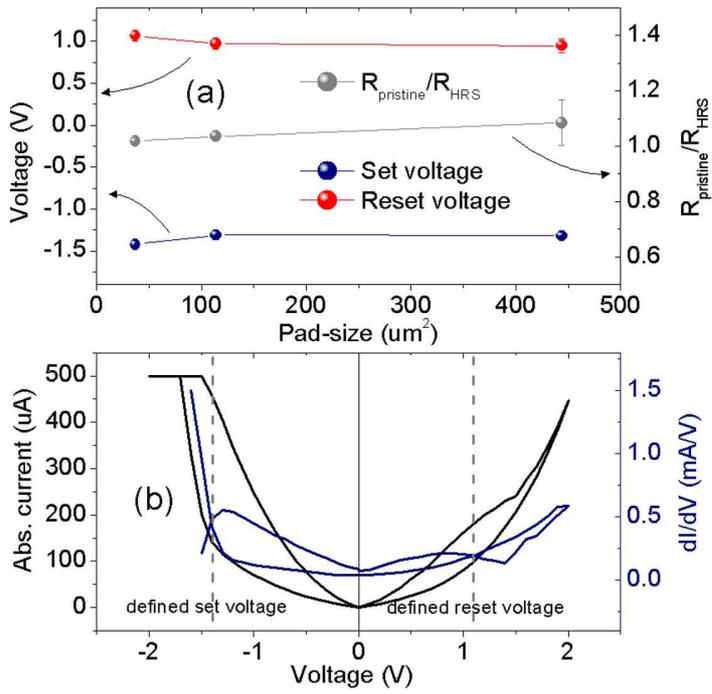

Figure 3

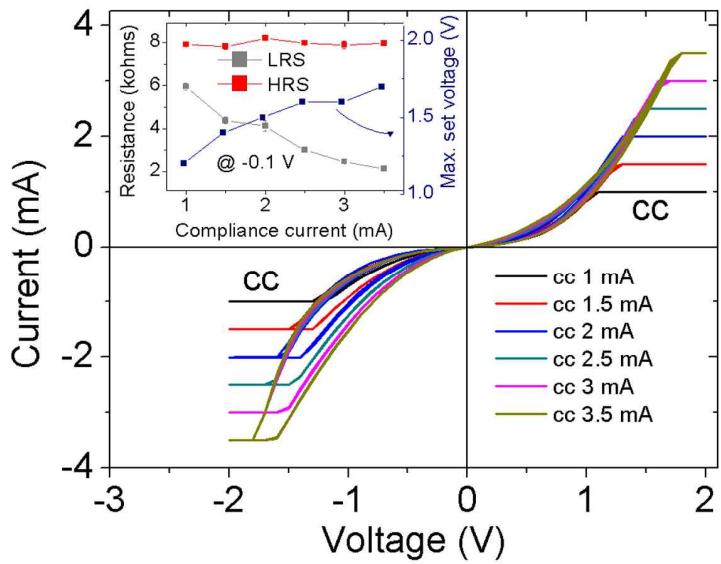

Figure 4